\documentclass[aip, reprint, floatfix]{revtex4-1}

\usepackage{dcolumn}
\usepackage{amsmath}
\usepackage{hhline}
\usepackage{float}
\usepackage{amssymb}
\usepackage{mathtools}
\usepackage{graphicx}
\usepackage[usenames, dvipsnames]{color}
\usepackage{bm}
\usepackage{url}
\usepackage[utf8]{inputenc}
\usepackage[T1]{fontenc}
\usepackage{mathptmx}
\usepackage{etoolbox}
\usepackage{subfigure}
\usepackage{slashed,bbm}
\usepackage{graphics,psfrag,epsfig}
\usepackage{dsfont}
\usepackage{setspace}
\usepackage{soul}
\usepackage{amsfonts}
\usepackage{siunitx}
\usepackage{geometry}
\usepackage{xr-hyper}
\usepackage{hyperref}
\usepackage{xcolor}
\usepackage{times}
\usepackage{verbatim}
\usepackage{newfile}

\definecolor{dark-red}{rgb}{0.4,0.15,0.15}
\definecolor{dark-blue}{rgb}{0.15,0.15,0.4}
\definecolor{medium-blue}{rgb}{0,0,0.5}
\hypersetup{colorlinks, linkcolor = {dark-blue}, citecolor = {dark-blue}, urlcolor = {medium-blue}}

\newcommand{\pdagger}{\phantom{\dagger}}

\makeatletter
\newcommand*{\toccontents}{\@starttoc{toc}}
\makeatother

\makeatletter
\def\@email#1#2{
 \endgroup
 \patchcmd{\titleblock@produce}
  {\frontmatter@RRAPformat}
  {\frontmatter@RRAPformat{\produce@RRAP{*#1\href{mailto:#2}{#2}}}\frontmatter@RRAPformat}
  {}{}
}
\makeatother

\begin{document}

\title{TMDs as a platform for spin liquid physics: \\ 
A strong coupling study of twisted bilayer WSe$_2$}

\author{Dominik Kiese}
\affiliation{Institute for Theoretical Physics, University of Cologne, 50937 Cologne, Germany}

\author{Yuchi He}
\affiliation{Institut f\"ur Theorie der Statistischen Physik, RWTH Aachen University and JARA-Fundamentals of Information Technology, 52056 Aachen, Germany}

\author{Ciar\'an Hickey}
\affiliation{Institute for Theoretical Physics, University of Cologne, 50937 Cologne, Germany}

\author{Angel Rubio}
\email{angel.rubio@mpsd.mpg.de}
\affiliation{Max Planck Institute for the Structure and Dynamics of Matter and Center for Free Electron Laser Science, Luruper Chaussee 149, 22761 Hamburg, Germany}
\affiliation{Center for Computational Quantum Physics, Simons Foundation Flatiron Institute, New York, NY 10010 USA}
\affiliation{Nano-Bio Spectroscopy Group,  Departamento de Fisica de Materiales, Universidad del Pa\'is Vasco, UPV/EHU- 20018 San Sebasti\'an, Spain}

\author{Dante M. Kennes}
\email{dante.kennes@rwth-aachen.de}
\affiliation{Institut f\"ur Theorie der Statistischen Physik, RWTH Aachen University and JARA-Fundamentals of Information Technology, 52056 Aachen, Germany}
\affiliation{Max Planck Institute for the Structure and Dynamics of Matter and Center for Free Electron Laser Science, Luruper Chaussee 149, 22761 Hamburg, Germany}

\date{\today}

\begin{abstract}
	The advent of twisted moir\'e heterostructures as a playground for strongly correlated electron physics has led to a plethora of experimental and theoretical efforts seeking to unravel the nature of the emergent superconducting and insulating states. Amongst these layered compositions of two dimensional materials, transition metal dichalcogenides (TMDs) are by now appreciated as highly-tunable platforms to simulate reinforced electronic interactions in the presence of low-energy bands with almost negligible bandwidth. Here, we focus on the twisted homobilayer WSe$_2$ and the insulating phase at half-filling of the flat bands reported therein. More specifically, we explore the possibility of realizing quantum spin liquid (QSL) physics on the basis of a strong coupling description, including up to second nearest neighbor Heisenberg couplings $J_1$ and $J_2$, as well as  Dzyaloshinskii-Moriya (DM) interactions. 
	Mapping out the global phase diagram as a function of an out-of-plane displacement field, we indeed find evidence for putative QSL states, albeit only close to SU$(2)$ symmetric points. In the presence of finite DM couplings and XXZ anisotropy, long-range order is predominantly present, with a mix of both commensurate and incommensurate magnetic phases.
\end{abstract}

\maketitle


\section{Introduction}

\begin{figure*}
    \centering 
    \includegraphics[width = 1.0\linewidth]{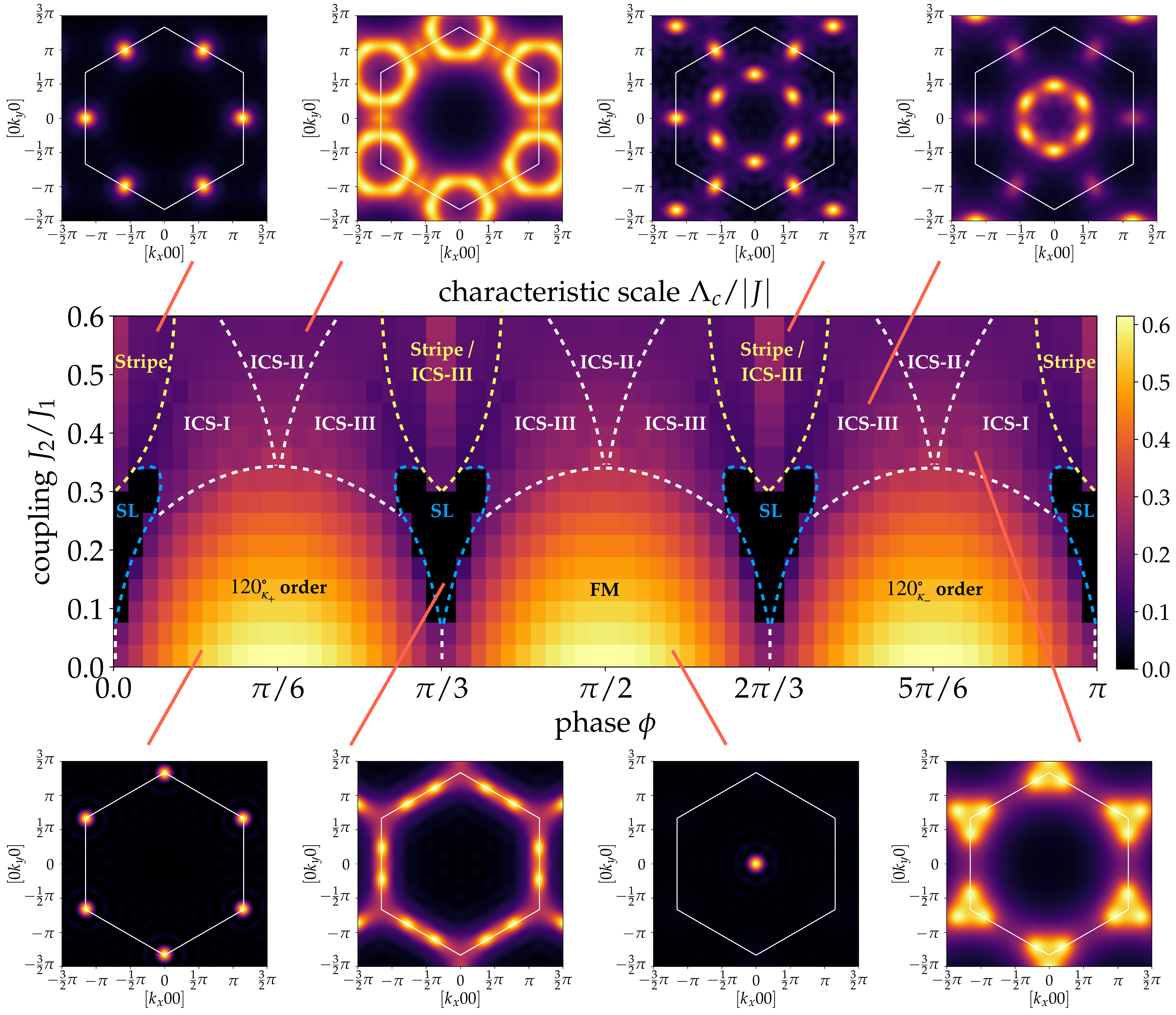}
    \caption{\textbf{Magnetic phase diagram for tWeS$_2$ obtained from pf-FRG.} We plot the characteristic RG scale $\Lambda_c$ indicating the emergence of magnetic long-range order or the absence thereof. In total, we identify a plethora of nine potential phases (SL = spin liquid, ICS = incommensurate spin spiral, FM = ferromagnet), including a putative quantum spin liquid for $\phi$ close to integer multiples of $\pi / 3$ and finite second-nearest neighbor Heisenberg coupling $J_2 / J_1$. The surrounding heat maps display the full elastic component of the structure factor (i.e. $\sum_{\mu} \chi^{\Lambda_c}_{\mu \mu}(\mathbf{k}, iw = 0)$), measurable, for example, by neutron scattering experiments. Further details about the different phases and how they are identified in our numerical calculations can be found in sections \ref{subsubsec:Classical}, \ref{subsubsec:FRG} and \ref{subsubsec:DMRG} of the main text.}
    \label{fig:FRG_global}
\end{figure*}

\noindent
Twisted moir\'e materials, such as the prominent magic-angle twisted bilayer graphene (tBG), have recently been established as a new platform to study many-body electron physics.\cite{Cao_2018_1, Cao_2018_2, Chen_2020, Liu_2020, KennesSimulator, Yoo_2019, balents_2020, chen_2019_1, chen_2019_2, shen_2020, cao_2020, burg_2019, he_2021, yankowitz_2019, lu_2019, stepanov_2020, sharpe_2019, serlin_2020, xu_2018, liu_2018, claassen2019universal, xian_2019, Kim_2020, Kerelsky_2019, Kerelskye2017366118, Halbertal2021, rubioverdu2020universal} The key mechanism promoting strongly enhanced electronic correlations is the formation of large moir\'e unit cells hosting low-energy bands with an extremely narrow bandwidth.\cite{Bistritzer12233, Mayou, CastroNeto} These flat bands have been shown to give rise to exotic low temperature phase diagrams featuring superconducting and insulating states, while offering a high degree of experimental control,\cite{KennesSimulator} e.g. over twist angle and doping. 

Recently, twisted bilayer transition metal dichalcogenides (tTMDs) have moved into the center of experimental attention as a tunable platform to simulate electronic many-body states.\cite{WangTransport,An_2020,Arora_2020, Xian_2021, TMD1, TMD2, Angeli_2021, naik_2018, regan_2020, Qian_2014, scherer2021mathcaln4} The decisive difference between tBG and tTMDs is the reduction of effective degrees of freedom in going from the former to the latter, allowing for the construction of simplified microscopic Hamiltonians, such as generalized Hubbard models, more amenable to (numerical) quantum many-body methods.\cite{TMD1, TMD2, tang_2020, Wu_2018, koshino_2018} 

Here, we consider a specific TMD bilayer, twisted WSe$_2$ (tWSe$_2$), for which a correlated insulating phase at half-filling of the flat bands has recently been reported.\cite{WangTransport, An_2020} These results have triggered corollary theoretical activity in deciphering the ground state phase diagram of the effective strong-coupling Hamiltonian,\cite{DasSarma, Millis} where the full rotation symmetry of the underlying triangular superlattice is broken down to $C_3$ by an anisotropic modulation of the spin couplings. The latter is parametrized by a phase $\phi$ inherited from the respective Hubbard model and can be tuned by an out-of-plane displacement field $V_{z}$. Notably, there is evidence from microscopic considerations,\cite{DasSarma} that large values of $|V_z| > 50$ meV support the emergence of second-nearest neighbor, SU$(2)$ symmetric Heisenberg exchange interactions. For the pure triangular lattice Heisenberg model, these are believed to undermine magnetic order in favor of a spin liquid ground state \cite{IqbalTriangular, HuTriangular, ZhuTriangular, Shimizu_2003, Szasz_2020} and as such, the intriguing possibility of realizing exotic phases in the exceptionally tunable experimental setup provided by twisted TMDs remains an interesting research direction. If experimentally realized, this would add elusive spin liquid states to the list of phases of matter accessible by controlled moir\'e engineering.\cite{KennesSimulator} 

In this manuscript, we set out to study the effective spin model proposed for tWSe$_2$, \cite{DasSarma, Millis} augmented by an antiferromagnetic second-nearest neighbor Heisenberg coupling $J_2$ previously not considered, using both classical \textit{and} quantum many-body methods. In the classical limit, we first use the Luttinger-Tisza method to determine the likely magnetic orders at zero temperature. We then investigate their stability with respect to thermal fluctuations and a strictly enforced constraint on the length of the classical $O(3)$ spins by performing classical Monte Carlo simulations. The quantum phase diagram is mapped out utilizing state-of-the-art pseudo-fermion functional renormalization group (pf-FRG) calculations and (infinite) density matrix renormalization group \cite{White, mcculloch2008infinite, Zauner_2015} techniques (iDMRG). 

Our key results are summarized in Fig.~\ref{fig:FRG_global}. In order to discuss them in a concise manner, we first focus on the regime $\phi \in [0, \frac{\pi} {6}]$, as the remainder of the phase diagram can be related via a simple three-sublattice mapping (see Sec. \ref{sec:Model}). 
The three main features of this regime can then be phrased in the following way: (1) Both classically and quantum mechanically we find that the $120^{\circ}$ order, featuring, for $\phi > 0$, a finite vector chirality $\kappa$ (discussed below), becomes more stable with increasing $\phi$. (2) At large $J_2$, finite $\phi$ tends to favor one of two incommensurate spin spiral states over the stripe order expected for the pure $J_1$-$J_2$ model. Classically, any finite $\phi$ suffices to generate incommensurate correlations, whereas quantum mechanically, the stripe order seems to remain stable for small $\phi$. (3) Close to the Heisenberg limit a paramagnetic region is identified for finite values of $J_2$, indicating a putative realm for quantum spin liquid physics. This regime, however, quickly diminishes with increasing $\phi$. These observations can straightforwardly be generalized to the parameter space beyond $\phi = \pi / 6$, albeit with new labels for the different phases. For example, close to $\phi = \pi / 3$ one finds a ferromagnetic ground state, instead of the chiral $120^{\circ}$ orders found at $\phi = 0$ and $\phi = 2\pi/3$.

The remainder of the manuscript is structured as follows. First, following the arguments of previous microscopic considerations,\cite{DasSarma} the derivation of the effective tWSe$_2$ spin model, starting from the corresponding tight-binding Hamiltonian, is recapped. We then summarize known results for the $J_2 = 0$ limit and elaborate on symmetry properties of the strong-coupling Hamiltonian. Second, the results obtained within the Luttinger-Tisza method and classical Monte Carlo simulations are discussed. Next, we introduce the pf-FRG and iDMRG methods and present their implications for the quantum phase diagram. We conclude by evaluating the relevance of our results for future experimental studies of tWSe$_2$ and pointing out further possible research directions.


\section{Model}
\label{sec:Model}

We focus on homobilayers of tWSe$_2$, which have recently been studied both experimentally,\cite{WangTransport,ZhangSTM} using  transport and scanning tunneling microscopy (STM) measurements, as well as theoretically,\cite{WangTransport,DasSarma, Millis} using mean-field approaches. The STM measurements have demonstrated that the moir\'e valence bands originate from the $\pm K$ valleys of the two TMD layers, while the $\Gamma$ valley is energetically disfavored. Spin degrees of freedom are thereby locked to one of the two valleys, giving rise to an effective spin-orbit coupling in the corresponding tight-binding Hamiltonian on the triangular superlattice \cite{DasSarma, Millis}
\begin{align}
    H_t = \sum_{\alpha \in \{ \uparrow, \downarrow \}} \sum_{ij} t^{\alpha}_{ij} c^{\dagger}_{i \alpha} c^{\phantom{\dagger}}_{j \alpha} + \text{h.~c.} \,.
    \label{eq:tb-ham}
\end{align}
Due to time-reversal and point group symmetries, the hoppings $t^{\alpha}_{ij}$ have to obey $t^{\alpha}_{ij} = \bar{t}^{\alpha}_{ji}$, $t^{\alpha}_{ij} = \bar{t}^{\bar{\alpha}}_{ij}$, while also being invariant under only threefold lattice rotational symmetry.\cite{DasSarma} This reduction from $C_6$ to $C_3$ results from changes in the Fermi surface topology when an external, out-of-plane displacement field is applied, which shifts the $K$ and $K'$ points of the mini-Brillouin zone in opposite directions.\cite{DasSarma} By combining the tight-binding Hamiltonian \eqref{eq:tb-ham} with an on-site interaction $U$, a spin-orbit coupled generalized Hubbard model results. This on-site interaction has been found to be about one order of magnitude larger than the kinetic contribution,\cite{DasSarma} motivating a strong-coupling description. 

\begin{figure}
    \centering 
    \includegraphics[width = 1.0\columnwidth]{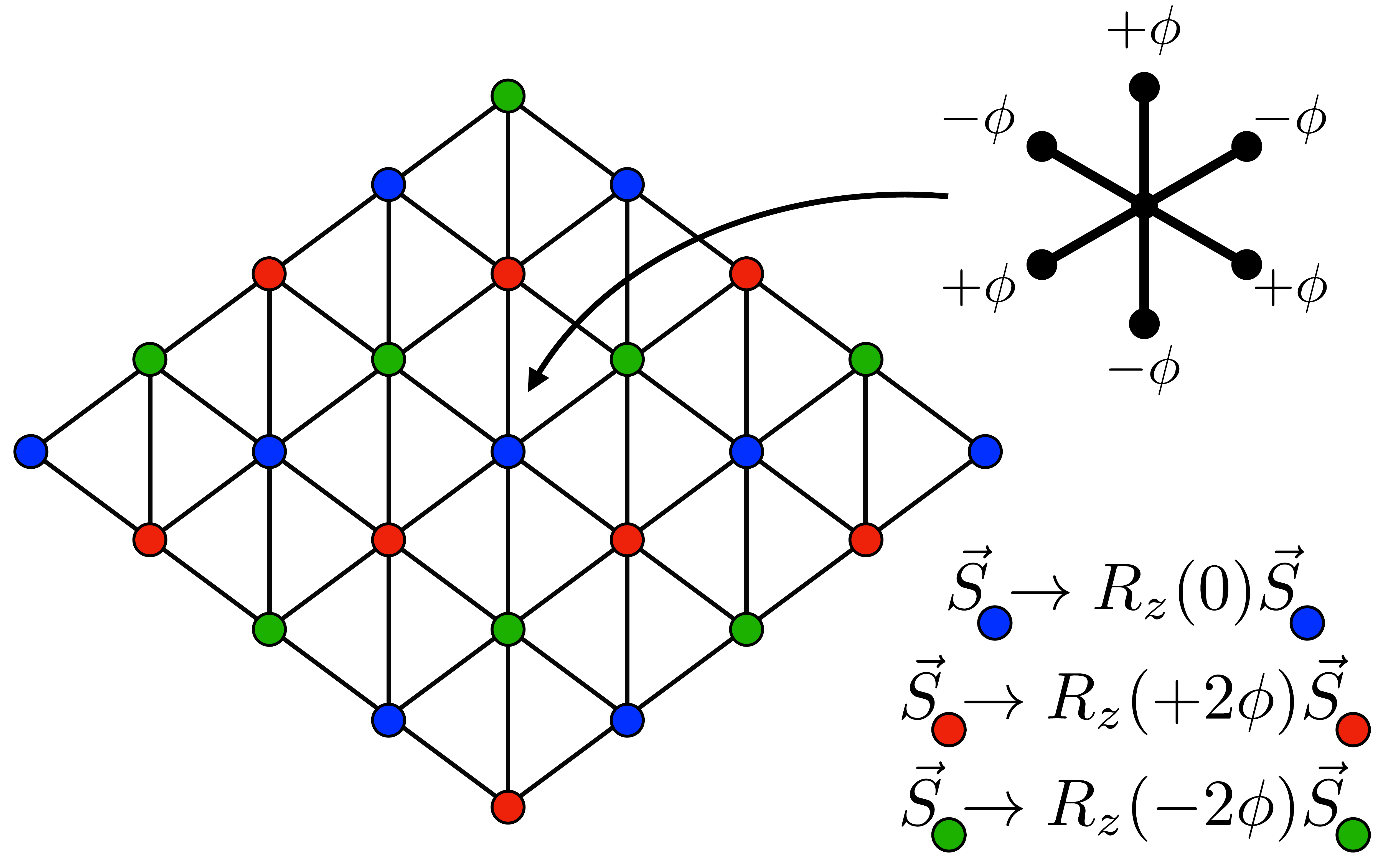}
    \caption{\textbf{Three sublattice rotation for the triangular lattice model.} The anisotropic phase $\phi_{ij}$ changes sign between nearest-neighbor bonds (as shown in the upper right corner). The Hamiltonian \eqref{eq:model} can be recast in terms of out-of-plane rotation matrices $R_{z}(-2\phi_{ij})$ (see Eq.~\eqref{eq:model_rotation}). By rotating the spins on the three sublattices (see lower right corner) each nearest-neighbor term in Eq.~\eqref{eq:model_rotation} can be transformed into an SU$(2)$ symmetric Heisenberg interaction, except for terms coupling the red and green sublattices. The remaining rotation, by $-6\phi$, vanishes for $\phi = n \frac{\pi}{3}$ with $n \in \mathbb{Z}$.}
    \label{fig:sublattice_trafo}
\end{figure}

In this work, we consider the $U \gg |t^{\alpha}_{ij}|$ limit at half-filling, where one can derive an effective spin model \cite{DasSarma, Millis} with residual U$(1)$ symmetry about the $z$-axis,
\begin{align}
    H =& \ J_{1} \sum_{\langle ij \rangle} \left[ \cos(2 \phi_{ij}) (S^{x}_{i} S^{x}_{j} + S^{y}_{i} S^{y}_{j}) + S^{z}_{i} S^{z}_{j}  \right] \notag \\
                &+ J_{1} \sum_{\langle ij \rangle}
                \sin(2 \phi_{ij}) \, \hat{z} \cdot \left( \mathbf{S}_i \times \mathbf{S}_j \right) \notag \\
                &+ J_{2} \sum_{\langle \langle ij \rangle \rangle} \mathbf{S}_i \cdot \mathbf{S}_j \,,
    \label{eq:model}
\end{align}
featuring XXZ, off-diagonal Dzyaloshinskii-Moriya (DM) and SU$(2)$ symmetric next-nearest neighbor Heisenberg interactions. The phase $\phi_{ij}$ varies sign between nearest-neighbor bonds (see  Fig.~\ref{fig:sublattice_trafo}), thus inheriting the reduction from sixfold to threefold lattice rotational symmetry from the tight-binding model \eqref{eq:tb-ham}. As pointed out in Ref.~\onlinecite{DasSarma}, the form of the underlying second-nearest neighbor hopping $t_2^\alpha$ motivates the inclusion of a fully SU$(2)$ symmetric Heisenberg interaction $J_2$, which has previously not been considered. For large displacement fields $|V_z| > 50 \ \text{meV}$, this $J_2$ is the next largest interaction beyond the nearest-neighbor $J_1$ terms. \cite{DasSarma}

For $J_2 = 0$, the ground state phase diagram of Eq.~\eqref{eq:model} has previously been studied using classical Luttinger-Tisza and self-consistent Hartree-Fock mean field calculations.\cite{DasSarma, Millis} For $\phi \in [0, \pi]$, both works find a ferromagnetic phase ($\pi / 3 < \phi < 2 \pi / 3$) sandwiched between two antiferromagnetic $120^{\circ}$ orders with opposite vector chiralities $\kappa_\pm$, with $\kappa = \text{sgn}(\hat{z} \cdot (\mathbf{S}_1 \times \mathbf{S}_2 + \mathbf{S}_2 \times \mathbf{S}_3 + \mathbf{S}_3 \times \mathbf{S}_2))$, where $\mathbf{S}_1, \mathbf{S}_2$ and  $\mathbf{S}_3$ are spins on a triangular plaquette. For finite $J_2$, however, the situation has not yet been studied and quantum fluctuations could stabilize more exotic phases, especially since several numerical works \cite{IqbalTriangular, HuTriangular, ZhuTriangular, Szasz_2020} suggest that the pure $J_1 - J_2$ Heisenberg model on the triangular lattice hosts a quantum spin liquid ground state. 

For finite $\phi = n \ \frac{\pi}{3}$ with $n \in \mathbb{Z}$, the nearest-neighbor terms can be transformed into a fully SU$(2)$ symmetric form by performing a three sublattice rotation (see Fig.~\ref{fig:sublattice_trafo}), thus opening up the possibility to experimentally tune the system close to the (effective) Heisenberg limit by variation of the displacement field.\cite{DasSarma, Millis} This can be clearly seen by rewriting the Hamiltonian as
\begin{align}
    H = J_{1} \sum_{\langle ij \rangle} \mathbf{S}_i^\textrm{T}  R_z(-2 \phi_{ij}) \mathbf{S}_j 
                + J_{2} \sum_{\langle \langle ij \rangle \rangle} \mathbf{S}_i \cdot \mathbf{S}_j \,,
    \label{eq:model_rotation}
\end{align}
where $R_z(-2 \phi_{ij})$ is an out-of-plane rotation matrix with rotation angle $-2 \phi_{ij}$, and then performing the transformation shown in Fig.~\ref{fig:sublattice_trafo}. Indeed, more generally, the energetics at $\phi$, $\phi+n\pi/3$, and $n\pi/3 - \phi$ are identical, though crucially the wavefunctions do change.  

With these observations in mind, we therefore focus our efforts on the regime $\phi \in [0, \frac{\pi}{6}]$ and study the respective ground states by classical Luttinger-Tisza and Monte Carlo simulations, as well as quantum pf-FRG and iDMRG calculations that go beyond mean-field theory. The global phase diagram can then be straightforwardly obtained using the sublattice rotation outlined above and adjusting the labels of the phases accordingly.


\section{Results}

\subsection{Classical Limit}
\label{subsubsec:Classical}

First we explore the classical $S\rightarrow\infty$ limit of the model. In order to determine the likely classical magnetic orders, we turn to the Luttinger-Tisza (LT) method. \cite{LT1946} This method treats the spin as an unconstrained vector, allowing for a straightforward Fourier transform and subsequent diagonalization of any quadratic spin Hamiltonian. For the model in Eq.~\eqref{eq:model}, the corresponding energy eigenvalues are

\begin{align}
    E_{H}(\mathbf{k}) =& \ J_1 \sum_{\boldsymbol{\delta}_{\textrm{1}}} \cos(\mathbf{k}\cdot \boldsymbol{\delta}_{\textrm{1}}) + J_2 \sum_{\boldsymbol{\delta}_{\textrm{2}}} \cos(\mathbf{k}\cdot \boldsymbol{\delta}_{\textrm{2}}) \notag \\
    E_{\pm}(\mathbf{k}) =& \ J_1 \sum_{\boldsymbol{\delta}_{\textrm{1}}} \cos(\mathbf{k}\cdot \boldsymbol{\delta}_{\textrm{1}} \pm 2\phi_{ij}) + J_2 \sum_{\boldsymbol{\delta}_{\textrm{2}}} \cos(\mathbf{k}\cdot \boldsymbol{\delta}_{\textrm{2}}) \,,
    \label{eq:LTevals}
\end{align}
where $\boldsymbol{\delta}_1$ and $\boldsymbol{\delta}_2$ are the set of nearest and next-nearest neighbor lattice vectors. $E_H (\mathbf{k})$ is independent of $\phi$ and is identical to the Heisenberg result (i.e.~$\phi=0$), with eigenvalue lying purely along the $z$-axis. On the other hand, $E_\pm (\mathbf{k})$ are explicitly $\phi$ dependent, with eigenvalues lying purely within the $xy$-plane. For a given set of parameters, the absolute minimum eigenvalue provides a strict lower bound to the classical energy, and the corresponding momentum, which we denote by $\mathbf{k}^\star$, provides a candidate classical ordering wavevector.  

For $\phi=0$, i.e.~the $J_1$-$J_2$ Heisenberg model, there is a transition from $120^\circ$ order, with ordering wavevector $\mathbf{k}^\star=K$, to stripe order, with $\mathbf{k}^\star=M$, at a critical value of $J_2/J_1 = 1/8$. Turning on a small finite $\phi\neq 0$ has three important consequences, (i) it forces the spins to order within the $xy$-plane ($E_{\pm}(\mathbf{k})$ is always favoured), (ii) it selects a definite chirality and helps stabilize the $120^\circ$ order, increasing its extent to a maximum of $J_2/J_1 = 1/3$ at $\phi=\pi/6$, and (iii) it immediately turns the stripe order incommensurate, which we label ICS-I, with an ordering wavevector $\mathbf{k}^\star$ that lies along the high-symmetry $M-K$ line (and $M-K'$ line, though from here on we will simply use $K$ when no further distinction is necessary). It also generates a new ordered phase, clustered close to $\phi=\pi/6$, with incommensurate magnetic order and an associated ordering wavevector that does not lie on any high-symmetry line, which we label ICS-II. 

\begin{figure*}
\centering
    \includegraphics[width=1.0\linewidth]{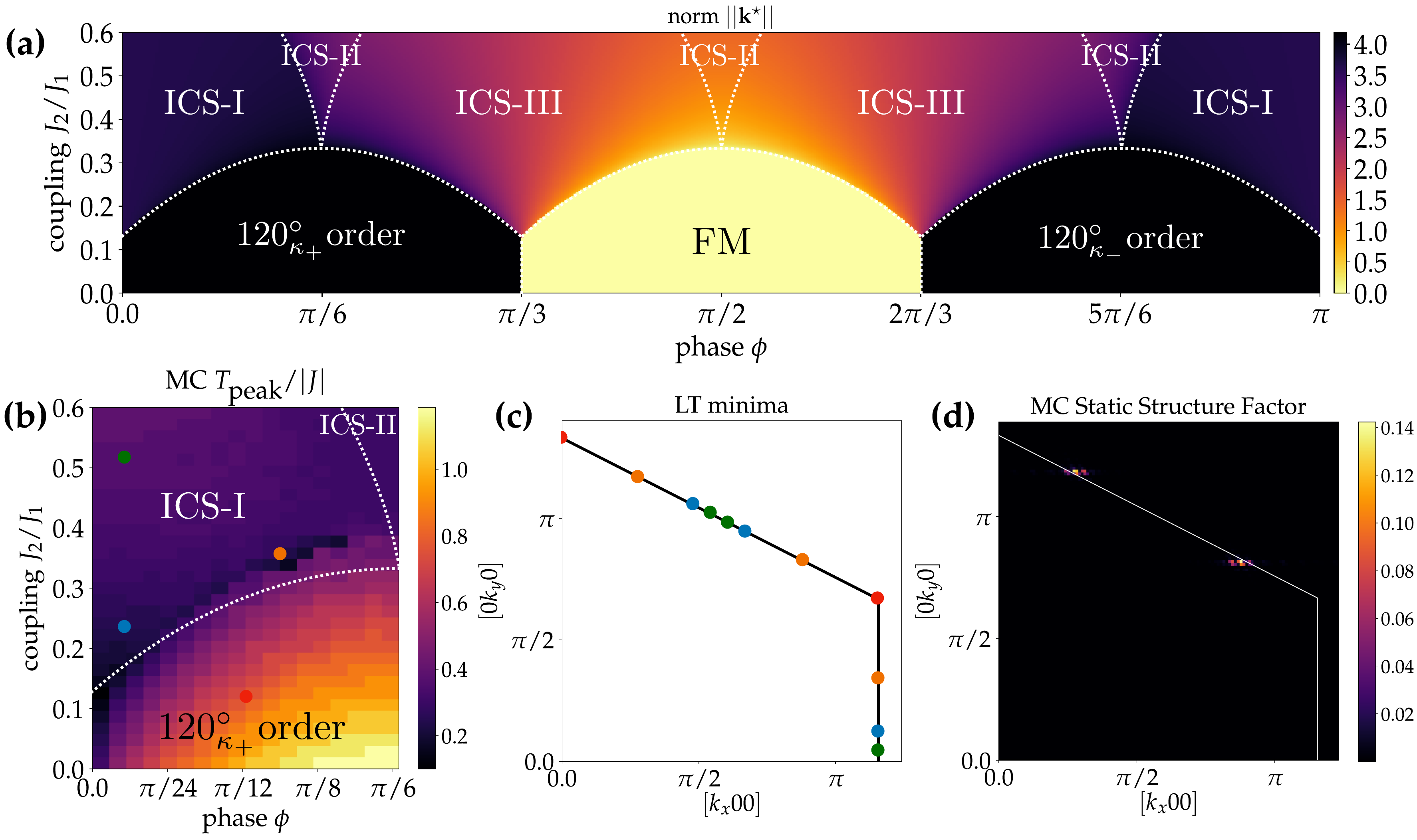}
    \caption{\textbf{Classical phase diagram} with (a) the Luttinger-Tisza (LT) result. Six distinct phases are visible, a simple ferromagnetic order (FM), two $120^\circ$ orders, with vector chirality $\kappa_+$ and $\kappa_-$, and three incommensurate phases, ICS-I, ICS-II and ICS-III. The background color indicates the norm of the ordering wavevector $||\mathbf{k}^\star||$. (b) Classical Monte Carlo results, on a $96\times 96$ lattice, for the peak temperature $T_{\text{peak}}$ from the specific heat across the phase diagram, with the LT phase boundaries overlaid on top. (c) Locations of the LT ordering wavevectors $\mathbf{k}^\star$ within the Brillouin zone for the four points marked in (b) (see Fig.~\ref{fig:FRG_phase_diagram} for the spin-$1/2$ pf-FRG structure factors at the same points). (d) Classical static structure factor at low temperature obtained by Monte Carlo simulations at the point $J_2/J_1=0.36$ and $\phi=\pi/10$ (orange point in (b) and (c)), within the ICS-I phase.}
    \label{fig:Classical_PD}
\end{figure*}

As noted in section \ref{sec:Model}, the physics of the model for $\phi>\pi/6$ can be related to the region $\phi \in [0, \frac{\pi} {6}]$ discussed above via a simple three-sublattice transformation. Indeed, this can also be seen from the form of the LT eigenvalues, with $E_{\pm}(\mathbf{k}) \rightarrow E_{\pm}(\mathbf{k} \pm n \mathbf{K}) $ for $\phi \rightarrow \phi + n\pi/3$. Thus, the $120^\circ$ order gets mapped to FM order, and the ICS-I phase, with $\mathbf{k}^\star$ along the $M-K$ line, gets mapped to a new ICS-III phase, with $\mathbf{k}^\star$ along the $K-\Gamma$ line. The classical phase diagram is summarised in Fig.~\ref{fig:Classical_PD}. 

Classical Monte Carlo (MC) simulations allow us to explore the relative stability of the different phases, as well as confirming that the LT predictions are correct. As the model contains a continuous U$(1)$ rotational symmetry about the $z$-axis, the Mermin-Wagner theorem precludes a finite in-plane magnetization at finite temperature. However, a peak in specific heat, at $T_{\textrm{peak}}$, related to a Berezinskii-Kosterlitz-Thouless (BKT) transition due to ordering in the $xy$-plane is still possible, \footnote{Note that, for a BKT transition, the specific heat $T_\textrm{peak}$ does not mark the location of the phase transition, but instead slightly overestimates it, e.g. $T_\textrm{peak} \sim 1.1-1.2 T_\textrm{BKT}$ for the triangular lattice XXZ model.\cite{Gupta1992,Capriotti1998}} as seen for example in the triangular lattice XXZ model (the first term in Eq.~\eqref{eq:model}). \cite{Capriotti1998} A map of $T_{\textrm{peak}}$ is shown in Fig.~\ref{fig:Classical_PD}(b) with, as expected, the highest $T_{\textrm{peak}}\sim 1.5 J_1$ occurring for $J_2/J_1=0$ and $\phi=\pi/6$, a consequence of the enhanced stability for the $120^\circ$ order that finite $\phi$ provides. On the other hand, within the incommensurate phases, $T_{\textrm{peak}}$ shows little variation, lying between $0.4-0.5 J_1$ for the whole range shown. Finally, Fig.~\ref{fig:Classical_PD}(d) shows an example of a MC static structure factor within the ICS-I phase whose peak precisely matches the prediction from LT. 

\subsection{Pseudo-Fermion Functional Renomalization Group}

\subsubsection{Method}
\label{subsubsec:FRG}

In the past decade, the pseudo-fermion functional renormalization group (pf-FRG) developed by Reuther and W\"olfle,\cite{ReutherOrig} has been widely employed to investigate ground state phase diagrams of quantum spin models on two \cite{ReutherOrig, Reuther-2011a} and three \cite{Iqbal3D} dimensional lattices. The method utilizes the parton decomposition
\begin{align}
    S^{\mu}_i = \frac{1}{2} \sum_{\alpha, \beta} f^{\dagger}_{i \alpha} \sigma^{\mu}_{\alpha \beta} f^{\pdagger}_{i \beta} \,,
    \label{eq:partons}
\end{align}
to recast the original Hamiltonian in terms of fermionic creation and annihilation operators. Here, $\sigma^{\mu}_{\alpha \beta}$ for $\mu \in \{x, y, z\}$ denote Pauli matrices. Changing the representation space of the spin algebra, however, comes with a caveat: The dimensions of the (local) Hilbert space of pseudo-fermions ($d = 4$) and spin-$1/2$ operators ($d = 2$) are different and as such the respective representations are not isomorphic. Although unphysical states can be eliminated by an additional local constraint $\sum_{\alpha} f^{\dagger}_{i\alpha} f^{\pdagger}_{i \alpha} = 1$ on every lattice site, an exact treatment of this constraint is rather difficult and in practice the softened condition $\langle \sum_{\alpha} f^{\dagger}_{i\alpha} f^{\pdagger}_{i \alpha} \rangle = 1$ is employed. Fluctuations around the mean have been found to leave observables computed within pf-FRG qualitatively unchanged,\cite{LargeS, thoenniss2020multiloop} advocating an on-average treatment of the fermionic number constraint at zero temperature. 

Having rewritten the spin Hamiltonian in terms of fermions, a regulator function, here chosen as
\begin{align}
    \Theta^{\Lambda}(w) = 1 -  e^{-w^2 / \Lambda^2} \,,
\end{align}
 with flow parameter $\Lambda$, is implemented in the bare propagator as
\begin{align}
    G_0(w) \rightarrow G^{\Lambda}_{0}(w) = \Theta^{\Lambda}(w) G_0(w) \,.
\end{align}
This procedure gives rise to $\Lambda$-dependent $n$-point correlation functions, whose flow from the ultraviolet $G^{\Lambda \to \infty}_{0}(w) = 0$ to the infrared $G^{\Lambda \to 0}_{0}(w) = G_0(w)$ limit is governed by a hierarchy of ordinary integro-differential flow equations. To be amenable to numerical algorithms, the latter has to be truncated. Here, we utilize the Katanin truncation, \cite{ReutherOrig, LargeN, LargeS} which cuts off the flow equations beyond the two-particle vertex and has been demonstrated to efficiently capture competing magnetic and non-magnetic phases. 
\footnote{Recently, an advanced, `multi-loop' truncation scheme has been successfully employed within the pf-FRG,\cite{thoenniss2020multiloop, kiese2021multiloop} improving on certain aspects of the Katanin truncation. Since the respective one and higher loop ground state phase diagrams were, however, qualitatively the same, we refrain from performing higher loop calculations in this work to reduce the numerical effort involved in treating the off-diagonal DM interaction in combination with breaking of $C_6$ symmetry.} 

The main observable extracted from the pf-FRG is the flowing spin-spin correlation function 
\begin{align}
	\chi^{\mu \nu \Lambda}_{ij}(iw = 0) = \int_{0}^{\beta} d\tau \langle T_{\tau} S^{\mu}_{i}(\tau) S^{\nu}_{j}(0) \rangle^{\Lambda} \,,
\end{align}
which shows an instability (like a cusp, kink or divergence) once the RG flow selects a ground state with broken symmetries. The absence of such a breakdown is consequently associated with paramagnetic phases such as spin liquids. Furthermore, for long-range ordered states, the respective type of magnetic order can be characterized by Fourier transforming $\chi^{\mu \nu \Lambda}_{ij}$ to momentum space ($F[\chi^{\mu \nu \Lambda}_{ij}](\mathbf{k}) = \chi^{\Lambda}_{\mu \nu}(\mathbf{k})$) and determining the wavevectors $\mathbf{k}_{\text{max}}$ with the largest spectral weight. Further information on the method and its numerical implementation are provided in Sec. I of the supplemental material. 

Due to the symmetry properties of Eq.~\eqref{eq:model}, we consider two distinct susceptibilities $\chi^{\Lambda}_{\text{XX}}(\mathbf{k})$ $(=\chi^{\Lambda}_{\text{YY}}(\mathbf{k}))$ and $\chi^{\Lambda}_{\text{ZZ}}(\mathbf{k})$ in momentum space to distinguish possible in-plane and out-of-plane magnetic orders. While finite in general for $\phi_{ij} > 0$, off-diagonal correlation functions $\chi^{\Lambda}_{\text{XY}}(\mathbf{k})$ $(= - \chi^{\Lambda}_{\text{YX}}(\mathbf{k}))$ turn out to be rather small compared to their diagonal counterparts in our pf-FRG calculations and are therefore only considered as a benchmark to check for a switch in vector chirality between the two $120^{\circ}$ orders.

\subsubsection{Phase diagram}

\begin{figure*}
    \centering 
    \includegraphics[width = 1.0\linewidth]{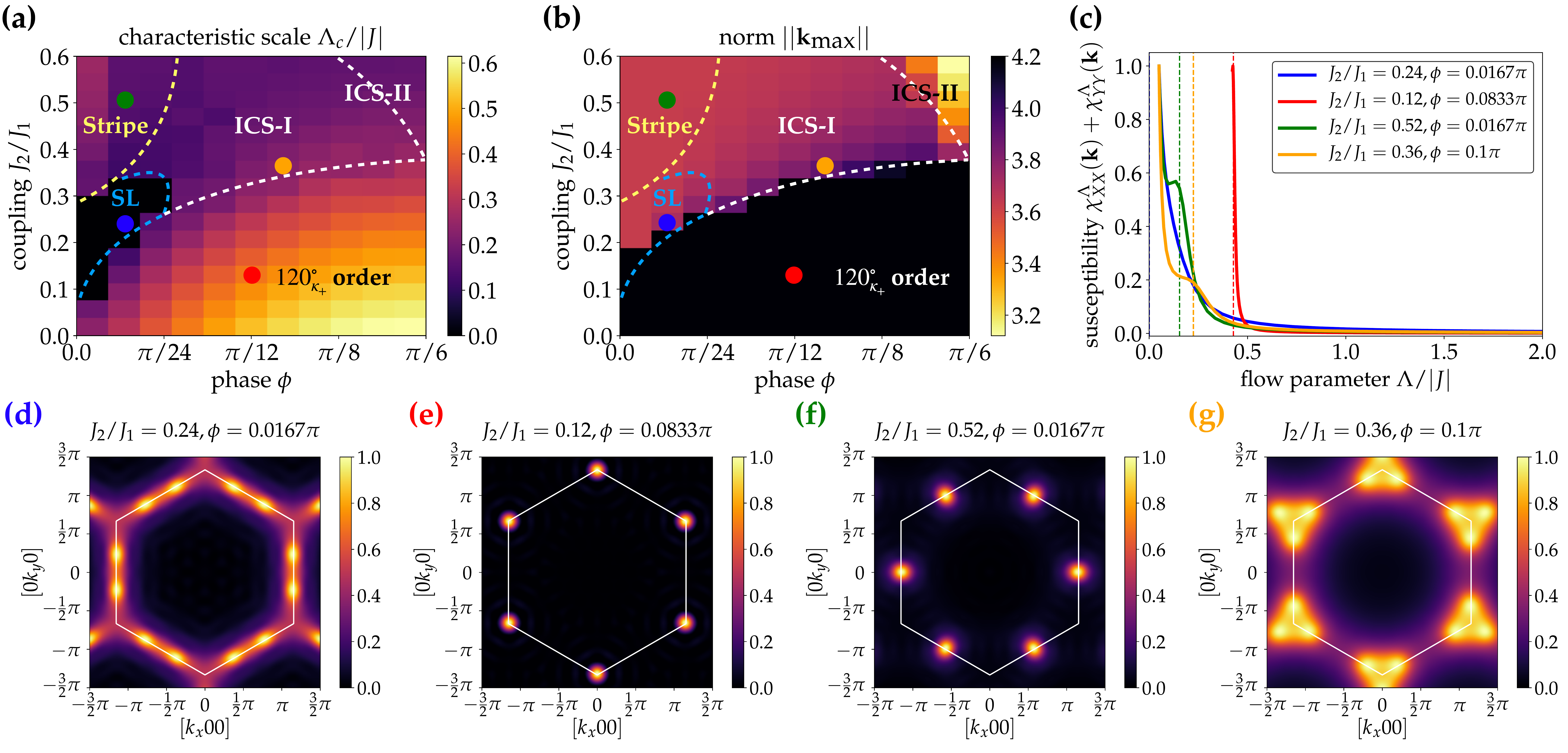}
    \caption{\textbf{Phase diagram for $\phi \in [0, \frac{\pi}{6}]$ obtained from pf-FRG.} In (a) the characteristic RG scale $\Lambda_c$ is shown as a function of antiferromagnetic next-nearest neighbor coupling $J_2 / J_1$ and phase $0 \leq \phi \leq \pi / 6$, with approximate phase boundaries drawn as a guide to the eye. We find a small region of quantum spin liquid (SL) behavior for small $\phi < \pi / 24$ and intermediate values of $0.12 < J_2 / J_1 < 0.32$, where the RG flow (see the blue curve in panel (c)) stays smooth and featureless down to the lowest accessible cutoff values. The rest of the phase diagram is occupied by four different magnetically ordered phases, which can be distinguished by their ordering wavevector $\mathbf{k}_{\text{max}}$ and its respective norm, as displayed in (b). For the stripe and $120^{\circ}$ ordered phases (with definite vector chirality $\kappa_+$), $\mathbf{k}_{\text{max}}$ resides at the $M$ and $K$ points respectively, whereas it continuously changes position in the spin liquid and incommensurate spin spiral (ICS) phases, as apparent from the color gradient in (b). In (c) we show representative flows of the magnetic susceptibility as a function of the RG scale $\Lambda / |J|$, with dashed lines highlighting the position of the characteristic scale $\Lambda_c / |J|$ (which is most visible for the in-plane correlators). The latter can be distinctly identified for the stripe and $120^{\circ}$ phase, whereas the incommensurate phases only show a pronounced shoulder, indicating strongly competing tendencies between magnetic and non-magnetic channels in the pf-FRG equations. The flows have been normalized by their respective maximum for better comparability. Finally, (d)-(g) display the full, diagonal structure factors $\sum_{\mu} \chi^{\Lambda}_{\mu \mu}(\mathbf{k})$ computed at the characteristic scale $\Lambda_c$ for the four points marked with colored dots in (a) and (b). 
    }
    \label{fig:FRG_phase_diagram}
\end{figure*}

We now turn to the discussion of the $\phi \in [0, \frac{\pi} {6}]$ region of the phase diagram of our model Hamiltonian Eq. \eqref{eq:model}, as obtained within pf-FRG and summarized in Fig.~\ref{fig:FRG_phase_diagram}. 

For small $\phi \lesssim \pi / 48$ and intermediate next nearest-neighbor coupling, we find a small region of spin liquid behavior, where the RG flow (see the blue curve in Fig.~\ref{fig:FRG_phase_diagram}(c)) stays smooth and featureless down to the lowest simulated cutoff value $\Lambda / |J| = 0.05$, where $|J|=\sqrt{J_1^2+J_2^2}$. For $\phi=0$, corresponding to the pure $J_1$-$J_2$ Heisenberg model, the estimated range of the spin liquid regime $0.12 \lesssim J_2 / J_1 \lesssim 0.32$ is larger than the respective literature values $0.06 - 0.08 \lesssim J_2 / J_1 \lesssim 0.15 - 0.17$, which we attribute to our softened treatment of the fermionic number constraint and the exclusion of higher loop corrections in the current framework. Since the FRG calculation is nevertheless capable of reproducing the existence of a paramagnetic regime between the adjacent $120^{\circ}$ and stripe ordered phases (consistent with previous studies \cite{Reuther-2011a}), we are confident that its qualitative predictions of the phase diagram are reliable. The structure factor $\sum_{\mu} \chi^{\Lambda}_{\mu \mu}(\mathbf{k})$ within the SL phase is displayed in Fig.~\ref{fig:FRG_phase_diagram}(d). It resembles an interpolation between the $120^{\circ}$ and stripe orders (Fig.~\ref{fig:FRG_phase_diagram}(e) \& (f)), in the sense that its peaks move on the high-symmetry line between the $K$ and $M$ points of the first Brillouin zone as $J_2$ and $\phi$ are increased. In this regard, the spin liquid region appears similar to a molten version of the neighboring incommensurate spin spiral phase (ICS-I in Fig.~\ref{fig:FRG_phase_diagram}(a) \& (b)), albeit with a washed out distribution of subleading weight along the Brillouin zone edges. The spectral weight for the ICS-I phase is, in contrast, much more localized, though of course the maxima still reside at incommensurate positions between the $K$ and $M$ points (Fig.~\ref{fig:FRG_phase_diagram}(g)). 

For larger $\phi$, we find the pf-FRG phase diagram to be roughly consistent with the classical result (Fig.~\ref{fig:Classical_PD}), predicting, for $J_2 / J_1 \gtrsim 0.32$, a transition from in-plane $120^{\circ}$ order to one of two incommensurate phases that can be distinctly identified by the position of their ordering wavevector $\mathbf{k}_{\text{max}}$ within the first Brillouin zone (Fig.~\ref{fig:FRG_phase_diagram}(b)). The phase boundary is, however, shifted upwards, in favor of the $120^{\circ}$ order, within the FRG. We generally find the dominant contributions to the structure factor to stem from the in-plane correlations, i.e. $\chi^{\Lambda}_{\text{XX}} + \chi^{\Lambda}_{\text{YY}}$, where flow breakdowns are most visible, though out-of-plane correlations become sizable with increasing $J_2$. This finding is in line with the Luttinger-Tisza result Eq.~\eqref{eq:LTevals}, as the eigenvalues corresponding to in-plane and out-of-plane order move closer together.

Notably, our pf-FRG approach also finds a stripe ordered ground state for $J_2 / J_1 \gtrsim 0.32 - 0.36$ and close to $\phi = 0$. In contrast, our classical calculations predict the stripe order to be unstable to  incommensurate ordering for any finite $\phi$. This could be, on the one hand side, due to finite-size effects in the pf-FRG calculations (though for increased lattice truncation ranges no changes are observed), which would make it difficult to decipher the extremely weak classical incommensuration at small $\phi$. On the other hand, quantum fluctuations may also favor the commensurate stripe order over the ICS-I phase, especially since their classical energies for small $\phi$ and large $J_2$ are almost degenerate. We also note that the signatures for magnetic ordering, as characterized by a breakdown of the RG flow (see Fig.~\ref{fig:FRG_phase_diagram}(c)) are rather weak in the incommensurate phases (pronounced shoulder versus sharp peak or divergence in the stripe and $120^{\circ}$ phase), hinting towards strongly competing magnetic and non-magnetic channels within the FRG approach for this part of the phase diagram. 

For $\phi > \pi / 6$, as before, the structure of the model outlined in Sec. \ref{sec:Model}, allow to us to straightforwardly generalize our results (see Fig.~\ref{fig:FRG_global}), while adjusting the labels for the different phases. Between $\pi / 3 < \phi < 2 \pi / 3$ antiferromagnetic $120^{\circ}$ order is replaced by a ferromagnetic ground state, which yet again becomes $120^{\circ}$ ordered, though with opposite vector chirality, for $2 \pi / 3 < \phi < \pi$. At large $J_2$ and $\pi / 6 < \phi < 5\pi /6$, the ICS-I order gets mapped to another incommensurate spin spiral phase (ICS-III) with susceptibility maxima located on the high-symmetry line between the $\Gamma$ and $K$ points. Finally, the stripe order found close to the Heisenberg limit $\phi = 0$ re-appears close to $\phi = \pi / 3$ and $\phi = 2 \pi / 3$ in coexistence with the ICS-III order (see Fig.~1 in the supplemental material for further details).

\subsection{Density Matrix Renormalization Group}
\label{subsubsec:DMRG}

\begin{figure*}
    \centering 
    \includegraphics[width = 1.0\linewidth]{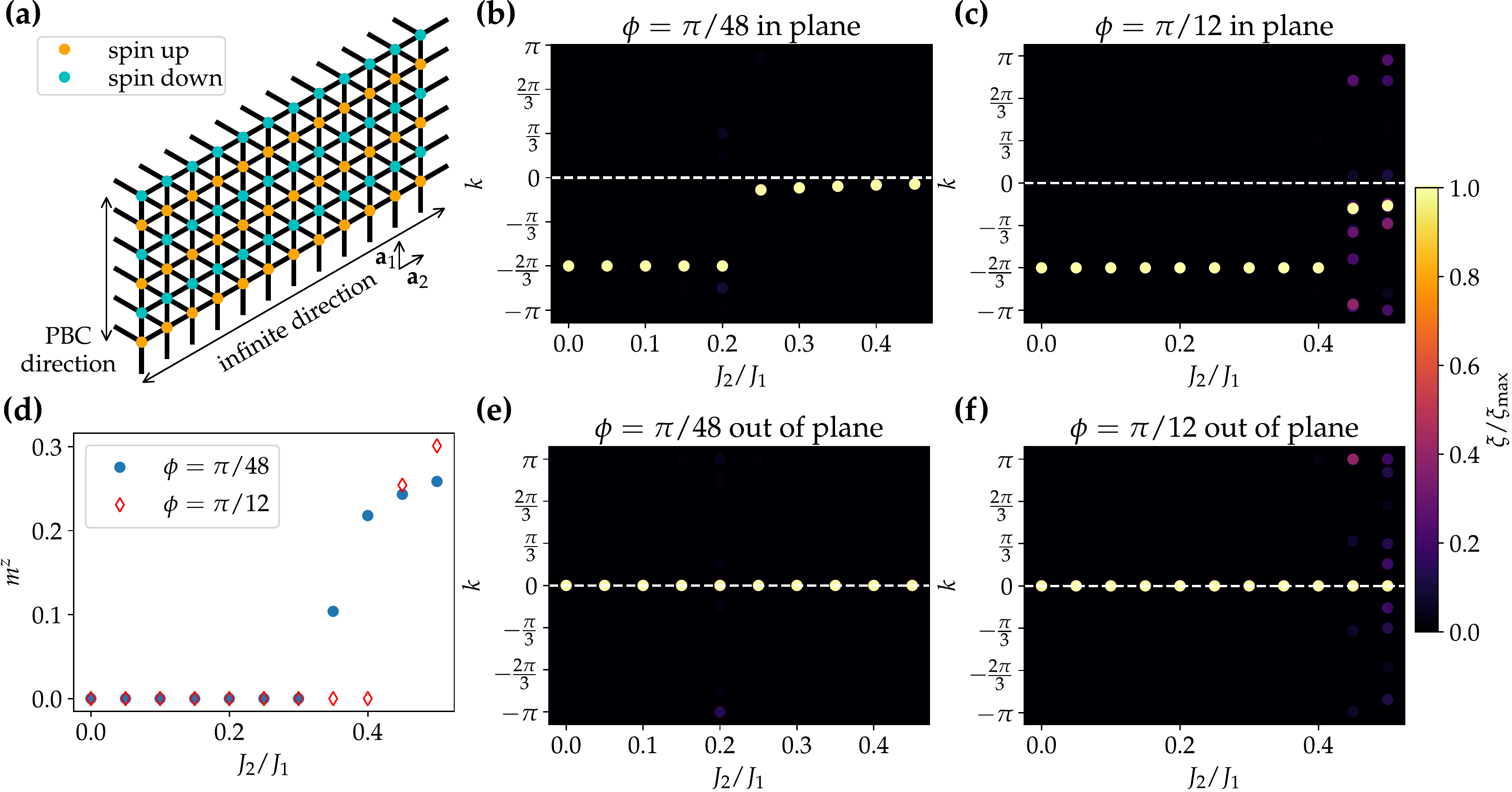}
    \caption{\textbf{Correlation length spectrum on an infinite cylinder geometry.} (a) shows the triangular lattice on an infinite cylinder geometry with an $L_y=6$ site circumference. A configuration of the possible out-of-plane stripe order is illustrated by the coloring of sites in orange (cyan), indicating an out-of-plane spin up (down). (b)-(f) show the correlation length spectra (see the main text for definition) along the infinite direction for $\phi=\pi/48$ and $\pi/12$. For each value of $J_2/J_1$, we plot the 20 largest correlation lengths at their respective $k$ values. For most cases, few points are visible as they share the same $k$. Finally, in (d), the out-of-plane staggered magnetization along the cylinder direction, $m^z$, is plotted. The evidence for the in-plane $120^{\circ}$ phase is the dominant in-plane correlation length at $-2\pi/3$. 
    In-plane incommensurate correlations are visible for relatively large $J_2$ in both (b) and (c), where the momenta are not locked to high symmetry points, but are instead distributed around $-\pi/10$ and $-\pi/6$ respectively. The indication for out-of-plane stripe order is given by a non-vanishing $m^z$ in (d).}
    \label{fig:iMPScorrelationandorder}
\end{figure*}

To complement our numerical results, we now present our iDMRG calculations of the model for two representative $J_2$ cuts at $\phi = \pi / 48$ and $\pi / 12$ on an infinite cylinder geometry. We use the two-site iDMRG algorithm \cite{White, mcculloch2008infinite} to optimize infinite matrix product states (iMPS) as approximations to the ground state wavefunctions. We chose the bond dimension such that the error is smaller than the marker size in every plot.~\footnote{We use bond dimension $3,600$ for Fig.~\ref{fig:iMPScorrelationandorder}(b),(c),(e), and (f), while using bond dimensions $3,600$ and $5,000$ for extrapolating~\cite{pinorder} the non-vanishing $m^{z}$ in (d).} The two-site iDMRG truncation errors are at most on the order of $10^{-7}$.

The cylinder geometry is illustrated in Fig.~\ref{fig:iMPScorrelationandorder}(a). We choose a circumference $L_y = 6$, compatible with the possible $120^{\circ}$ and stripe orders, with an example of the latter shown in the same subfigure.
The infinite cylinder geometry then allows us to probe possible incommensurate correlations along the infinite direction. 


Recall that finite $\phi$ explicitly breaks SU$(2)$ symmetry down to a residual in-plane U$(1)$ symmetry. According to the Mermin-Wagner theorem for quasi-one dimensional systems (such as our cylindrical iDMRG geometry), an in-plane $120^{\circ}$ order that spontaneously breaks U$(1)$ symmetry is forbidden. However, the existence of a possible two-dimensional $120^{\circ}$ phase can be inferred by studying spin-spin correlations. On the other hand, long-range out-of-plane stripe order does not break any continuous symmetry and can therefore be directly observed within our iDMRG calculations.

 
We study the out-of-plane, 
$\left\langle S^{z}_0 S^{z}_{n\mathbf{a}_2} \right\rangle$, and in-plane, 
$\left\langle S^{+}_0 S^{-}_{n\mathbf{a}_2} \right\rangle$, spin-spin correlation functions, where $\mathbf{a}_2$ is the lattice vector along the infinite direction (see Fig.~\ref{fig:iMPScorrelationandorder}(a)) and $S_i^{\pm}=S_i^{x}\pm iS_i^{y}$. 
Using the iMPS data, it is known that static correlation functions of this form can be written as $\sum_{j} C_{j} e^{ik_{j}n}e^{-n/\xi_j} \ $, \cite{Zauner_2015} where $j$ sums over eigenvectors of the iMPS transfer matrix. The largest $\xi_j$ corresponds to the dominant correlation length, while the respective $k_j$ then characterizes the momentum of the lowest-lying excitation along the infinite direction. 
The correlation length spectrum has, for example, been used to study the $\phi=0$ case in Ref.~\onlinecite{HuTriangular}.  

Within the correlation length spectrum, the $120^{\circ}$ order for the SU$(2)$ symmetric case corresponds to a dominant correlation length at $k = \pm 2\pi/3$, equal in magnitude for both in- and out-of-plane components.
For finite $\phi$, however, our classical and pf-FRG calculations indicate that $120^{\circ}$ order is not only locked to the $xy$-plane but also locked to a certain chirality. 
Our DMRG data (see Fig.~\ref{fig:iMPScorrelationandorder}) is consistent with these results, as we observe that only in-plane correlations display a peak at $k =-2 \pi / 3$.


For large $J_2$ and $\phi \neq 0$, we find incommensurate correlations characterized by a continuously varying momentum in Figs.~\ref{fig:iMPScorrelationandorder}(b) \& (c). Curiously, the incommensurate correlations can exist either with or without an accompanying finite out-of-plane staggered magnetization along the cylinder direction, $m^z =  \sum_y |\langle S^z_y e^{i\pi y} \rangle| /L_y$. For the chosen cylinder geometry, an out-of-plane stripe order, with stripes parallel to the infinite $\mathbf{a}_2$ direction (shown in Fig.~\ref{fig:iMPScorrelationandorder}(a)), has a finite $m^z$. For $\phi = \pi / 48$, we observe a relatively large region, $0.2 \lesssim J_2/J_1 \lesssim 0.3$, with in-plane incommensurate correlations and with negligible out-of-plane components, $S_i^z\approx 0$. This is consistent with the pf-FRG structure factor computed in the putative spin liquid phase, Fig.~\ref{fig:FRG_phase_diagram}(d), where residual but broadened incommensurate peaks are visible. On the other hand, for larger $J_2$, we obtain a finite out-of-plane $m^z$ (see Fig.~\ref{fig:iMPScorrelationandorder}(d)), consistent with the onset of out-of-plane stripe order. For $\phi = \pi / 12$, we observe however, at least within our resolution, just a single direct transition from in-plane $120^\circ$ order to out-of-plane stripe order, not inconsistent with the absence of a spin liquid in the pf-FRG calculations. 

Note, however, that at $\phi=\pi/12$ the pf-FRG predicts that out-of-plane stripe order is much weaker compared to the in-plane incommensurate correlations (see Fig.~\ref{fig:FRG_phase_diagram}(e), where peaks at the $M$ point, coming from the out-of-plane component of the structure factor, are overshadowed by the in-plane incommensurate peaks). The stripe order that we identify in iDMRG may be a finite size artifact of the quasi one-dimensional cylinder geometry, with the possibility that stripe order is molten in favor of the incommensurate in-plane order when going to two dimensions. As the incommensurate correlations are frustrated along the finite direction of the cylinder, the finite size effects should in fact be rather large. Our iDMRG calculations may in turn be biased toward commensurate stripe order, as opposed to an incommensurate phase. 
Further simulations with larger $L_y$, beyond the scope of this work, are necessary to settle on a final conclusion regarding this issue.

\section{Discussion}
Twisted TMDs have been predicted to provide an exciting opportunity to realize the physics of the triangular lattice Hubbard model, and potentially access the magnetism of its strong-coupling limit. By focusing on the particular case of tWSe$_2$, and including both first- and second-nearest neighbor couplings, as well as a finite displacement field, we have mapped out the strong coupling phase diagram. Perhaps the most intriguing phase, the QSL that appears in the pure $J_1$-$J_2$ limit, unfortunately only inhabits a small portion of the larger phase diagram, which includes XXZ anisotropy and an effective DM interaction. Accessing QSL physics thus requires tuning the displacement field such that $\phi\sim n \pi/3$, and the twist angle such that $J_2/J_1$ is within the required range. It is an open question whether further interactions, generated by taking into account further hoppings $t_{ij}^\alpha$ and interactions $U_{ij}$ of the underlying Hubbard model, can lead to a wider, more stable QSL window.  

A large part of the phase diagram, above a sufficiently large $J_2/J_1 \sim 0.3$, hosts incommensurate magnetic phases. Such phases can be expected to host gapless phason modes, due to the low-energy cost of translating the incommensurate magnetic structure. This is on top of the underlying moir\'e structure, which, at the atomic level, is generically incommensurate. If it's possible to tune to such a large $J_2/J_1$ ratio, it would allow to explore the interplay between the moir\'e scale incommensurate magnetic structure, and its gapless phason modes, with the atomic scale incommensurate lattice structure, and its gapless phonon modes. \cite{Koshino2019,Maity2020}    

For smaller ratios of $J_2/J_1 \lesssim 0.3$, the $120^\circ$ order is stabilized. For $\phi=\pi/6$ it is particularly stable, and has a fixed chirality, which leaves only a single BKT transition at finite temperature, with an expected $T_{\textrm{BKT}} \gtrsim J_1$. It thus provides a particularly clean example of BKT physics within potential experimental reach, and the possibility of exploring moir\'e scale magnetic vortices.   

An important additional tuning parameter to consider in the future is an external magnetic field. It's effects on the $120^\circ$ order and $J_1$-$J_2$ QSL are already known, \cite{Ye2017,Nakano2017} but how it will distort the incommensurate phases found at finite $\phi$ is not immediately clear. An interesting possibility would be the formation of multi-Q states. Indeed, such a possibility is actually realised for incommensurate phases found within the pure $J_1$-$J_2$ Heisenberg model. \cite{Okubo2012} In that case, it is even possible to stabilize a skyrmion lattice phase at finite temperature. Realising a similar scenario for the model at hand, with incommensurate phases ICS-I, II and III, would open up a path to studying moir\'e scale skyrmion lattices within tWSe$_2$.\cite{Akram_2021} 

The phase diagram uncovered in this work expands our view on the landscape of opportunities arising within tTMDs. In particular, the strong coupling physics of tWSe$_2$ has the potential to realize and tune between QSLs, incommensurate magnetic orders and extremely stable, chiral $120^\circ$ and ferromagnetic orders. Adding QSL states and incommensurate magnetic orders to the catalogue of moir\'e-controllable phases of matter is an exciting open experimental question, which might be in reach using highly-tunable TMDs. We note that the case of tWSe$_2$ was taken here as a prominent experimentally characterized homobilayer example, but the available range of TMDs might help to fabricate other twisted van der Waals materials. In those, e.g., the QSL state could  take a more prominent stage in the respective  phase diagram.   

\textit{Note added}: During the completion of this manuscript we became aware of the publication of related (but previously inaccessible) work by Zare and Mosadeq.\cite{Honeycomb} In contrast to our study they focus on a honeycomb lattice model, rather than a triangular lattice model, which is then analyzed using the Luttinger-Tisza method (combined with a variational approach
to optimize the classical ground states) and DMRG simulations. They find similar conclusions regarding the fate of the quantum spin liquid phases and the stability of magnetic orders.

\acknowledgments
We thank M. Claassen, M. M. Scherer and Zhenyue Zhu for useful discussions. D.K. thanks L. Gresista and T. M\"uller for related work on the \texttt{PFFRGSolver.jl}
package \cite{kiese2021multiloop} used for the FRG calculations. The DMRG calculations are based on the Tenpy package \cite{10.21468/SciPostPhysLectNotes.5}. D.K. and C.H. acknowledge support from the Deutsche Forschungsgemeinschaft (DFG, German Research Foundation), Project No.~277146847, SFB 1238 (Project No.~C03). Y.H. and D.M.K. acknowledges funding by the Deutsche Forschungsgemeinschaft (DFG, German Research Foundation) under RTG 1995, within the Priority Program SPP 2244 ``2DMP'' and within Germany’s Excellence Strategy - Cluster of Excellence Matter and Light for Quantum Computing (ML4Q) EXC 2004/1 - 390534769. This work was supported by the Max Planck-New York City Center for Nonequilibrium Quantum Phenomena.
The numerical simulations were performed on the CHEOPS cluster at RRZK Cologne, the JURECA Booster \cite{jureca} and JUWELS cluster \cite{JUWELS} at the Forschungszentrum Juelich and the Raven cluster at MPCDF of the Max Planck society. 


\bibliography{bib_tWSe2}


\end{document}


\title{Supplemental Material: \\
\large \textit{TMDs as a platform for spin liquid physics: \\ 
A strong coupling study of twisted bilayer WSe$_2$}}

\maketitle
\onecolumngrid

\section{Pseudo-Fermion Functional Renomalization Group}
\label{supp:FRG}

\noindent
In this section, further technical details of the pf-FRG approach are discussed. As already mentioned in the main text, the hierarchy of differential equations at the heart of the FRG method has to be truncated to allow for a numerical solution. Within the Katanin truncation, the single-scale propagator $S^{\Lambda} \equiv -\diff G^{\Lambda}|_{\Sigma^{\Lambda} = \text{const.}}$ is replaced by a full derivative of the dressed propagator $G^{\Lambda}$ to partially include certain diagrammatic contributions of the three-particle vertex in the two-particle vertex flow. The flow equations for the self energy $\Sigma^{\Lambda}$ and the two-particle vertex $\Gamma^{\Lambda}$ then read
\begin{align}
	\diff \Sigma^{\Lambda}(1) &= -\frac{1}{2\pi} \sum_2 \G{1}{2}{1}{2} S^{\Lambda}(2)
	\label{supp:eq:self}
\end{align}
\begin{align}
	\diff &\G{1'}{2'}{1}{2} = \frac{1}{2\pi} \sum_{3, 4} \big{[} \G{3}{4}{1}{2} \G{1'}{2'}{3}{4}         \notag  \\ 
	&- \G{1'}{4}{1}{3} \G{3}{2'}{4}{2} - (3 \leftrightarrow 4)                  \notag  \\ 
	&+ \G{2'}{4}{1}{3} \G{3}{1'}{4}{2} + (3 \leftrightarrow 4) \big{]}          \notag  \\ 
	&\times G^{\Lambda}(3) \bigg{(} -\frac{d}{d\Lambda} G^{\Lambda}(4) \bigg{)} \,,
	\label{supp:eq:2Pgeneral}
\end{align}
where multi-indices comprise a lattice, spin and frequency argument, e.g. $1 = (i_1, \alpha_1, w_1)$. The flow equations can be further simplified by exploiting symmetries in real and spin space, as well as in the Matsubara frequencies. Since these simplifications are extensively discussed in Ref.~\onlinecite{BuessenOffdiag}, we only state the most important results here. Firstly, for time-reversal symmetric Hamiltonians, all one particle objects are diagonal in their indices and only depend on one frequency argument. Note, that this property has already been used to simplify \eqref{supp:eq:self} and \eqref{supp:eq:2Pgeneral} and the self energy, for example, should be regarded as
\begin{align}
    \Sigma(1) = \Sigma(w_1)
\end{align}
where $\Sigma(w) = -i\gamma(w)$ is purely imaginary and anti-symmetric in frequency space. The two particle vertex, on the other hand, is a bi-local object with purely real and purely imaginary components that encode the spin interaction of the respective Hamiltonian. For our model, we may write
\begin{align}
    \Gamma(1', 2'; 1, 2) &= \bigg{[} \Gamma^{\text{XX}}_{i_{1'}i_{2'}}(w_{1'}, w_{2'}, w_{1}) \times (\sigma^{x}_{\alpha_{1'} \alpha_1} \sigma^{x}_{\alpha_{2'} \alpha_2} + \sigma^{y}_{\alpha_{1'} \alpha_1} \sigma^{y}_{\alpha_{2'} \alpha_2}) \notag \\
    &+ \Gamma^{\text{ZZ}}_{i_{1'}i_{2'}}(w_{1'}, w_{2'}, w_{1}) \times \sigma^{z}_{\alpha_{1'} \alpha_1} \sigma^{z}_{\alpha_{2'} \alpha_2} \notag \\
    &+ \Gamma^{\text{DM}}_{i_{1'}i_{2'}}(w_{1'}, w_{2'}, w_{1}) \times (\sigma^{x}_{\alpha_{1'} \alpha_1} \sigma^{y}_{\alpha_{2'} \alpha_2} - \sigma^{y}_{\alpha_{1'} \alpha_1} \sigma^{x}_{\alpha_{2'} \alpha_2}) \notag \\
    &+ \Gamma^{\text{DD}}_{i_{1'}i_{2'}}(w_{1'}, w_{2'}, w_{1}) \times \delta_{\alpha_{1'} \alpha_1} \delta_{\alpha_{2'} \alpha_2} \notag \\
    &+ i\Gamma^{\text{ZD}}_{i_{1'}i_{2'}}(w_{1'}, w_{2'}, w_{1}) \times \sigma^{z}_{\alpha_{1'} \alpha_1} \delta_{\alpha_{2'} \alpha_2} \notag \\
    &+ i\Gamma^{\text{DZ}}_{i_{1'}i_{2'}}(w_{1'}, w_{2'}, w_{1}) \times \delta_{\alpha_{1'} \alpha_1} \sigma^{z}_{\alpha_{2'} \alpha_2} \bigg{]} \times \delta_{i_{1'}i_1} \delta_{i_{2'}i_2} \delta(w_{1'} + w_{2'} - w_1 - w_2) - (1' \leftrightarrow 2') \,.
\end{align}
The initial conditions for the flow equations then read
\begin{align}
    \gamma^{\Lambda \to \infty}(w) &= 0 \notag \\ 
    \Gamma^{\text{XX} \ \Lambda \to \infty}_{i_{1'}i_{2'}}(w_{1'}, w_{2'}, w_{1}) &= \frac{J_1}{4} \text{cos}(2\phi_{i_{1'}i_{2'}}) \times \mathbf{1}_{\langle i_{1'}i_{2'} \rangle} + \frac{J_2}{4} \times \mathbf{1}_{\langle \langle i_{1'}i_{2'} \rangle \rangle} \notag \\
    \Gamma^{\text{ZZ} \ \Lambda \to \infty}_{i_{1'}i_{2'}}(w_{1'}, w_{2'}, w_{1}) &= \frac{J_1}{4} \times \mathbf{1}_{\langle i_{1'}i_{2'} \rangle} + \frac{J_2}{4} \times \mathbf{1}_{\langle \langle i_{1'}i_{2'} \rangle \rangle} \notag \\
    \Gamma^{\text{DM} \ \Lambda \to \infty}_{i_{1'}i_{2'}}(w_{1'}, w_{2'}, w_{1}) &= \frac{J_1}{4} \text{sin}(2\phi_{i_{1'}i_{2'}}) \times \mathbf{1}_{\langle i_{1'}i_{2'} \rangle} \notag \\
    \Gamma^{\text{DD} \ \Lambda \to \infty}_{i_{1'}i_{2'}}(w_{1'}, w_{2'}, w_{1}) &= 0 \notag \\
    \Gamma^{\text{ZD} \ \Lambda \to \infty}_{i_{1'}i_{2'}}(w_{1'}, w_{2'}, w_{1}) &= 0 \notag \\
    \Gamma^{\text{DZ} \ \Lambda \to \infty}_{i_{1'}i_{2'}}(w_{1'}, w_{2'}, w_{1}) &= 0 \,,
\end{align}
where, for example, $\mathbf{1}_{\langle ij \rangle}$ is should be understood as
\begin{align}
    \mathbf{1}_{\langle ij \rangle} =
    \begin{cases}
        1\text{  if i and j are nearest-neighbors} \\
        0\text{  else}
    \end{cases}
\end{align}
For our work, we extend the open-source Julia package \texttt{PFFRGSolver.jl} \cite{kiese2021multiloop}, which provides a state-of-the-art pf-FRG solver for various lattice structures. We use a set of $N_{\Sigma} = 500$  frequencies for the self energy and $N_{\Gamma} = 50 \times 60$ transfer/fermionic frequencies to model the two-particle vertex. The lattice truncation is fixed to $L = 10$, i.e. correlations are set to zero beyond $10$ bonds away from a given reference site. The accuracy for the involved integration routines has been set to $a_{\text{tol}}, r_{\text{tol}} = (10^{-5}, 10^{-3})$ for which our results were found to be well converged. \\
In principle, the RG flow should diverge, once symmetries of the Hamiltonian are spontaneously broken during the flow to strong coupling. However, due to the relaxed particle number constraint and finite numerical resolution, phases with strongly competing channels occasionally develop softened features such as pronounced shoulders. Reading off the precise value of the characteristic scale $\Lambda_c$ in this case is rather difficult and a numerical criterion is needed to automate this process. Here, we utilize that the bare susceptibility $\chi^{\Lambda}_{0}(w = 0) = \int_{-\infty}^{\infty} dv (G^{\Lambda}_{0}(v))^2 \sim 1/\Lambda$, where $G^{\Lambda}_{0}(v) = (1 - e^{-v^2 / \Lambda^2}) / (iv)$ is the regularized bare propagator. Now, whenever the flow shows a distinct divergence or a sharp peak, we set $\Lambda_c$ to the position of this respective feature. If the breakdown is more washed out, we instead determine the scale with the strongest concavity, i.e. the largest deviation from the expected behavior of the non-interacting, paramagnetic system. If none of these criteria apply and the flow remains convex and featureless, we classify the ground state as non-magnetic.

\section{pf-FRG results for $\phi > \pi / 6$}
\label{supp:FRG_data}

\begin{figure}
    \centering 
    \includegraphics[width = 1.0\linewidth]{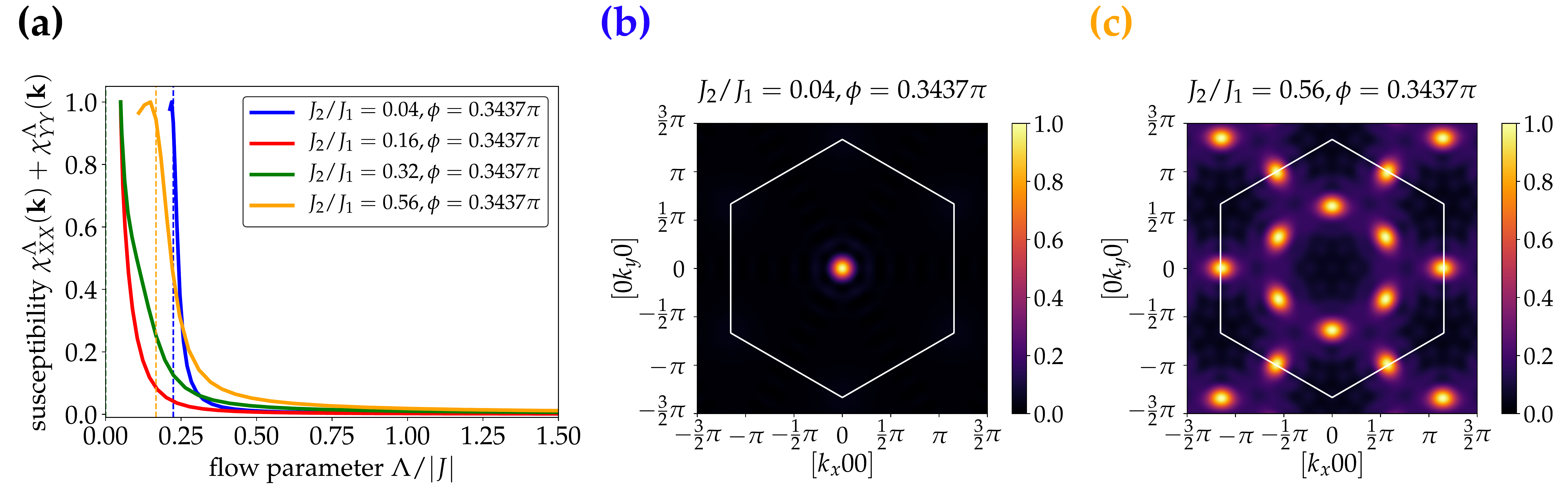}
    \caption{\textbf{pf-FRG data at $\pi/3 + \pi/96$.} In panel (a) we display representative RG flows, indicating, as expected from the considerations in Sec. II of the main text, a paramagnetic phase, quenched between the long-range ordered ferromagnetic (b) and stripe / type-III incommensurate phases (c). The incommensurate part of the correlation spectrum is generated by the in-plane correlations, where out-of-plane spin correlators $\chi^{\Lambda}_{\text{ZZ}}$ contribute the peaks at the $M$ points (related to stripy long-range order).}
    \label{supp:fig:FRG_further}
\end{figure}

In this section we discuss numerical results beyond the $[0, \frac{\pi}{6}]$ range focused on in the main text. As discussed in Sec. II of the main text, one would expect the energy spectrum of our Hamiltonian to be repeated within periods of $\pi / 3$. While the ground state wave function, and therefore the label of the respective phase, will generally change, the spin liquid region for which the characteristic energy scale $\Lambda_c$ vanishes, should re-appear upon varying $\phi.$ To illustrate this circumstance we have summarized representative pf-FRG data for $\phi = \pi/3 + \pi/96$ in Fig.~\ref{supp:fig:FRG_further}. Indeed, smooth RG flows, indicating a paramagnetic ground state, are obtained within the expected $J_2$ range, accompanied by sharp flow breakdowns in the adjacent magnetic phases. For large $J_2$, our FRG approach implies the presence of another stripy state in coexistence with the nearby ICS-III phase as visible from Fig.~\ref{supp:fig:FRG_further}(c), where Bragg peaks for both orders are visible. The incommensurate correlations are contributed by the in-plane, that is $\chi^{\Lambda}_{\text{XX} / \text{YY}}$, components of the susceptibility, whereas the peaks at the M-points steam from  $\chi^{\Lambda}_{\text{ZZ}}$.

\onecolumngrid

\bibliography{bib_tWSe2}